# Towards Precision Feeding Using Behavioral Monitoring in Marine Cages*

Dimitra Georgopoulou, Charalabos Vouidaskis, and Nikos Papandroulakis

*Abstract*— Aquaculture is expected to account for two-thirds of global fish consumption by 2030, highlighting the need for sustainable and efficient practices. Feeding is crucial to aquaculture success, influenced by factors like fish size, environment, and health. This study addresses a gap in feeding control for sea cages by developing a real-time monitoring system, using AI models and computer vision to analyze feeding behavior with European sea bass as pilot species. Key metrics like fish speed and a new feeding behavior index (FBI) were used to assess responses to different feeding scenarios. The results revealed distinct behavior patterns based on feeding quantity, with imbalances in activity when fish are overfed or underfed. The results can be used for predicting the level of satiation of the fish and controlling feeding duration.

## I. INTRODUCTION

Efficient feeding in aquaculture is crucial for fish growth, welfare, and environmental sustainability. Optimal feeding requires monitoring both behavioral and water quality parameters, as factors like temperature, oxygen, and pH significantly impact fish metabolism and appetite [1,2,3,4]. Research shows that activity levels and spatial positioning are linked to hunger and feeding patterns [5,6]. Hungrier fish exhibit more exploration and aggression [7], while well-fed fish are less active. Behavioral changes indicate appetite and feeding needs [8,9], and thus behavior monitoring can help maximize feed efficiency, prevent overfeeding, reduce stress from underfeeding, lower costs, and improve fish welfare. These insights have led to advanced monitoring systems, from visual observations to automated methods like computer vision and acoustic telemetry, enabling precise, non-invasive feeding control in aquaculture systems.

Despite advancements, there are significant gaps between experimental results and large-scale, commercial applications. For example, most research has focused on controlled environments like recirculating aquaculture systems (RAS), leaving sea cages, which host larger fish populations and experience fluctuating environmental conditions, underexplored. To address this gap, we have developed a continuous, real-time monitoring system for sea cages using European sea bass (*Dicentrarchus labrax*) as pilot species. This system utilizes underwater cameras and AI models to track individual fish and analyze their feeding behavior, introducing control parameters such as fish speed and the newly introduced feeding behavior index (FBI). By evaluating these indicators across different feeding scenarios, this study aims to enhance feeding efficiency in sea cages and contribute to more sustainable aquaculture practices.

## II. METHODOLOGY

### A. Experimental Animals and Site

A group of over 10,000 European sea bass (220 ± 30 g) was reared at a stocking density of 5.2 kg/m³ in a 40m diameter, 9m deep circular polyester cage at the HCMR pilot farm in Souda Bay, Crete. The cage had a cylinder-shaped net up to 8m deep with a 1m closing cone, and the fish were transferred to the farm as juveniles (approx. 2g) from the HCMR hatchery after 120 days post-hatching. An automatic feeder, controlled by a Raspberry Pi microcomputer, was used to test fish behavior under different feeding quantities: normal (based on feeding tables), reduced (50%), overfeeding (150%), and no feeding.

### B. Video Analysis

Swimming behavior analysis was conducted at both individual and group levels (Fig. 1). For individual fish detection and tracking, we trained YOLOv5 [10] on 1,000 annotated images and applied the DEEPSORT algorithm for tracking [11], excluding appearance-based association. Group-level analysis involved using computer vision to extract parameters like the feeding behavior index (FBI), which was based on group density variation during feeding. All video and data analysis, including speed calculations, were performed using Python (v3.9) on a desktop with an Intel Core i7-8700 CPU, 32GB RAM, and NVIDIA GeForce 3060Ti GPU.

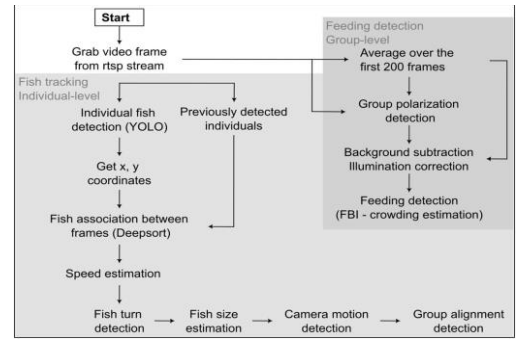

Figure 1: Flowchart of the algorithm.

### C. Data Analysis

The methodology for analyzing fish behavior included the following key steps:

- **Polynomial Fit:** A polynomial curve of degree 10 was fitted to the speed and Feeding Behavior Index (FBI) data to reduce noise, allowing for smoother and more accurate analysis.

* Research supported by the EU Horizon 2020 iFishIENCi project (818036).

Dimitra Georgopoulou is with the Institute of Marine Biology, Biotechnology and Aquaculture, Hellenic Center for Marine Research, AquaLabs, Thalassocosmos, Heraklion, Greece (corresponding author to provide phone: +302810337779; e-mail: d.georgopoulou@hcmr.gr).

Charalabos Vouidaskis was with Institute of Marine Biology, Biotechnology and Aquaculture, Hellenic Center for Marine Research, AquaLabs, Thalassocosmos, Heraklion, Greece (e-mail: mpampis.vouidaskis@gmail.com).

Nikos Papandroulakis is with the Institute of Marine Biology, Biotechnology and Aquaculture, Hellenic Center for Marine Research, AquaLabs, Thalassocosmos, Heraklion, Greece (e-mail: npap@hcmr.gr).

- **Asymmetry Calculation:** Changes in fish activity before and after feeding were compared using an asymmetry parameter. Positive values indicated greater activity after feeding.
- **Duration of Excitation:** The time during which fish activity exceeded the average post-feeding was calculated and normalized by the total feeding time for comparison across experiments.
- **Statistical Analysis:** Non-parametric Kruskal-Wallis tests and post hoc Dunn tests were applied to assess statistical differences in speed, FBI, and asymmetry across varying feeding quantities.
- **FBI Clustering:** The FBI signal was grouped into four clusters using Gaussian Mixture Models (GMM) to identify temporal changes in feeding behavior.

## III. RESULTS

Feeding quantity significantly impacted both fish speed and the Feeding Behavior Index (FBI), as shown in Fig. 2.

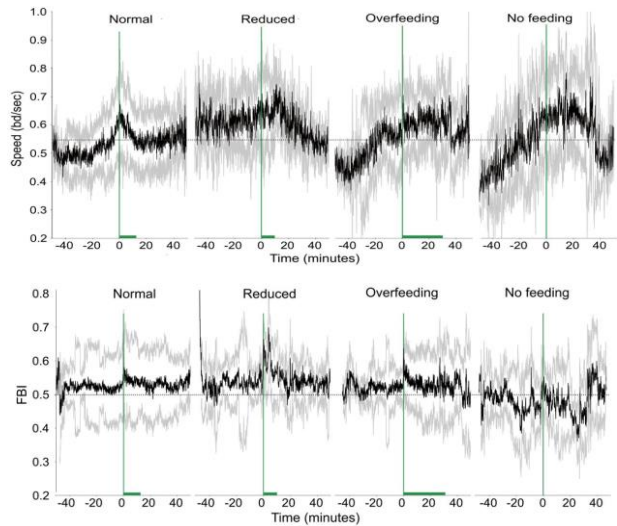

Figure 2: Speed and FBI in time for different feeding quantities. The green dashed vertical line shows when feeding starts and the green horizontal bar the duration of feeding. The dashed horizontal lines are reference lines to help comparison between the different scenarios.

During normal feeding, speed increased from 0.5 to 0.6 bd/sec (mean 0.54 ± 0.11 bd/sec) before feeding, then gradually decreased. In reduced feeding, activity remained elevated longer, with higher speeds (0.61 ± 0.1 bd/sec) compared to normal feeding (Statistic = 4.48, p-value = 0.03), though not significantly different from overfeeding (0.57 ± 0.12 bd/sec) or no feeding (0.56 ± 0.15 bd/sec). Overfeeding caused increased activity during feeding, while in no feeding, activity spiked around the expected feeding time. No significant differences were found in the duration of excitation and the average asymmetry between conditions (Statistic = 1.57; p-value = 0.67).

The FBI changes around feeding times vary across feeding scenarios (Fig. 2). In all cases, fish responded immediately to feeding with a sharp FBI increase. The increase was sharper, and the FBI signal was larger during reduced feeding compared to normal feeding, while it was smaller during overfeeding. Despite these observed differences, statistical analysis did not show significant differences for either the raw FBI (Statistic = 5.66, p-value = 0.13) or FBI asymmetry (Statistic = 5.67, p-value = 0.13).

Using the normal feeding scenario as a reference, distinct FBI clusters were identified using the Gaussian Mixture Model, resulting in four clusters: pre-feeding (black), two feeding clusters (red and grey), and post-feeding (blue) (Fig. 3). The first feeding cluster (red) occurs immediately after feeding starts, lasting nearly half of the feeding period, while the second (grey) continues until feeding ends. During overfeeding, the post-feeding cluster appears 10 minutes before feeding stops, while in reduced feeding, the second cluster extends beyond feeding. During fasting, the pre-feeding cluster is prolonged, partially covering the expected feeding period.

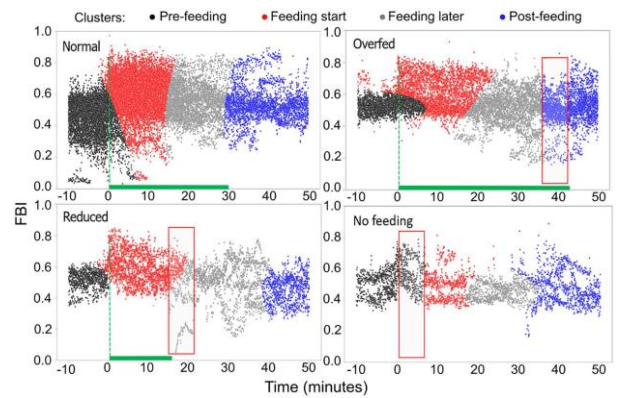

Figure 3: Clustering of the FBI for different feeding quantities (raw data). Black color: the pre-feeding, red color: feeding cluster immediately after feeding starts, grey color: feeding cluster at the later stages of feeding, blue color: post-feeding cluster.

## IV. DISCUSSION

In the current work, we developed a system for continuous monitoring of European sea bass feeding behavior in sea cages using AI models (YOLO, DEEPSORT) and computer vision techniques. A long-term experiment with varying feeding scenarios revealed that fish exhibit distinct behavioral patterns, measured through group speed and the feeding behavior index (FBI), which can help identify satiation levels and facilitate feeding control. We identified four FBI clusters—pre-feeding, two feeding phases, and post-feeding—showing how feeding quantities influence their duration. While full feeding control hasn't been achieved yet, our system provides valuable tools and reference curves for managing fish feeding, with potential applications for other species like gilthead seabream and salmon. Our current aim is to use changepoint analysis and neural networks to predict the abovementioned behavioral parameters and future feeding needs, aiding real-time control. Further research is needed to refine the system and explore additional feeding scenarios.

## REFERENCES

[1] D. L. Kramer, "Dissolved oxygen and fish behavior," in *Environ. Biol. Fishes* 18 (2), 81–92, 1987. doi: 10.1007/BF00002597

[2] C. Zhou, D. Xu, K. Lin, C. Sun, X Yang. "Intelligent feeding control methods in aquaculture with an emphasis on fish: a review," in *Rev. Aquac,* 10, 975–993, 2018, doi: 10.1111/raq.12218

[3] H. Volkoff, and I. Rønnestad, "Effects of temperature on feeding and digestive processes in fish," *Temperature* 7.4, 2020, 307-320.

[4] D. Assan, Y. Huang, U. F. Mustapha, M. N. Addah, G. Li, H. Chen, "Fish feed intake, feeding behavior, and the physiological response of apelin to fasting and refeeding," *Frontiers in endocrinology* 12, 2021, 798903.

[5] M. J. Hansen, T. M. Schaerf, A. J. W. Ward, "The effect of hunger on the exploratory behaviour of shoals of mosquitofish Gambusia holbrooki," *Behaviour* 152, 1659–1677, 2015, doi: 10.1163/1568539X-00003298

[6] T. Miyazaki, R. Masuda, S. Furuta, K. Tsukamoto, "Feeding behaviour of hatchery-reared juveniles of the Japanese flounder following a period of starvation," 2000, doi: 10.1016/S0044-8486(00)00385-9

[7] F. Huntingford, P. Tamilselvan, H. Jenjan, "Why do some fish fight more than others?," *Physiol. Biochem, Zoology* 85, 584–593, *2012,* doi: 10.1086/668204

[8] I. H. Chen, D. G. Georgopoulou, L. O. E. Ebbesson, D. Voskakis, P. Lal, N. Papandroulakis, "Food anticipatory behaviour on European seabass in sea cages: activity-, positioning-, and density-based approaches," Frontiers in Marine Science 10, 2023, 1168953.

[9] M. Conde-Sieira, M. Chivite, J. M. Míguez, J. L. Soengas, "Stress effects on the mechanisms regulating appetite in teleost fish," Front. Endocrinol, 2018, (Lausanne) 9, doi: 10.3389/fendo.2018.00631

[10] G. Jocher, A. Chaurasia, A. Stoken, J. Borovec, Y. Kwon, K. Michael, et al., "ultralytics/yolov5: v7.0 - YOLOv5 SOTA realtime instance segmentation (Zenodo)", 2022, doi: 10.5281/zenodo.7347926

[11] N. Wojke, A. Bewley, D. Paulus, "Simple online and realtime tracking with a deep association metric," in IEEE International Conference on Image Processing (ICIP), 2017, 3645–3649.